# Astro2020 APC White Paper

# Collaboration with Integrity:
# Indigenous Knowledge in 21st Century Astronomy

**Thematic Area:** State of the Profession Considerations

**Principal Author and Contact Information:** Aparna Venkatesan, University of San Francisco, avenkatesan@usfca.edu

**Co-Authors (alphabetical order):**

David Begay, Indigenous Education Institute, and University of Washington, dbegay@gmail.com
Adam Burgasser, U. of California San Diego, aburgasser@ucsd.edu
Isabel Hawkins, San Francisco Exploratorium, ihawkins@exploratorium.edu
Ka'iu Kimura, 'Imiloa Astronomy Center of Hawai'i, kaiukimura@gmail.com
Nancy Maryboy, Indigenous Education Institute, and University of Washington, dilyehe@gmail.com
Laura Peticolas, Sonoma State University, laurap@universe.sonoma.edu
Gregory Rudnick, University of Kansas, grudnick@ku.edu
Doug Simons, Canada France Hawai'i Telescope, simons@cfht.hawaii.edu
Sarah Tuttle, University of Washington, tuttlese@uw.edu

**Endorsers:**

Kimberly Coble, San Francisco State University, kcoble@sfsu.edu
Dara Norman, NOAO, dnorman@noao.edu

**Executive Summary:**

As the oldest science common to all human cultures, astronomy has a unique connection to indigenous knowledge (IK) and the long history of indigenous scientific contributions. Many STEM disciplines, agencies and institutions have begun to do the work of recruiting and retaining underrepresented minorities, including indigenous, Native American and Native Hawai'ian professionals. However, with the expansion of telescope facilities on sacred tribal or indigenous lands in recent decades, and the current urgency of global crises related to climate, food/water sovereignty and the future of humanity, science and astronomy have the opportunity more than ever to partner with indigenous communities and respect the wealth of sustainable practices and solutions inherently present in IK. We share a number of highly successful current initiatives that point the way to a successful model of "collaboration with integrity" between western and indigenous scholars. Such models deserve serious consideration for sustained funding at local and institutional levels. We also share six key recommendations for funding agencies that we believe will be important first steps for non-indigenous institutions to fully dialog and partner with indigenous communities and IK to build together towards a more inclusive, sustainable and empowering scientific enterprise.

**Introduction:**

Science is inseparable from the people conducting the science. Numerous white papers in the APC category of Astro2020 will address issues related to equity and inclusion, and propose targeted solutions to increase the gender, race and identity diversity in astronomy (see, e.g. papers by Norman et al., Zellner et al., and Burgasser et al.). These solutions aim for more than merely increasing diversity in the future workforce of astronomy. The research from the past few decades is clear that diverse teams work harder, provoke more thought, and result in higher-quality scientific research (Philips 2014). Groups of diverse problem solvers can outperform high-ability groups that lack problem-solving diversity (Hong & Page 2004). Simply put, diversity catalyzes creativity and innovation, markedly more so than homogeneity.

Among the diversity goals for astronomy, the chronic underrepresentation of indigenous people in STEM fields is a longstanding issue that persists despite numerous local and national efforts to address this. It is however becoming increasingly evident that there is an interconnection between astronomical scientific and educational activities and Indigenous Knowledge (IK). From the historical understanding of the cosmos and unique perspectives of time and space originally perceived by indigenous peoples, to conflicts over the expansion of telescope facilities on sacred tribal or indigenous lands, it has become more important than ever to respect, dialog, and partner with indigenous people and communities in our scientific endeavors. The richness of astronomical traditions from various indigenous cultures can be juxtaposed with western astronomy in a way that enhances science education and research while honoring the integrity and authenticity of indigenous perspectives (Venkatesan & Burgasser 2017). Traditional IK concepts tied to Earth and sky, developing "Collaboration with Integrity" (Maryboy, Begay & Peticolas 2012), deep listening, and sustainability, are relevant for both improved communication of astronomical research with the broader community and for better understanding of community tensions over proposed astronomical facilities.

In this white paper, we share specific initiatives pioneered by the Indigenous Education Institute[1], the 'Imiloa Astronomy Center in Hawai'i[2], and I-WISE[3] (Indigenous Worldviews in Informal Science Education) that demonstrate successful examples of cocreating educational and professional development experiences that involve understanding and embracing the deep historical, cultural and traditional language base of IK. These highly successful initiatives provided outcomes that demonstrate how the future vitality and capability of the astronomy and astrophysics workforce can be greatly strengthened by honoring IK, while helping to create a scientific enterprise that is more sustainable and inclusive. Such efforts have generated a successful model of collaboration with integrity between western and indigenous scholars that deserves serious consideration for sustained funding at local and institutional levels.

**We present six specific actionable recommendations for the funding agencies, drawn from our collective experience of being professional astronomers and knowledge-holders in a variety of formal and informal astronomy research/learning environments.** These recommendations are first steps towards developing sustainable inclusive practices in astronomy, and with the support of funding agencies, they can be put into practice before the next decadal survey of astronomy and astrophysics in 2030. This will move us closer to co-creating a modern scientific enterprise that includes the great wealth of IK across the numerous scientific subfields already contained holistically in IK.

---

[1] Indigenous Education Institute (IEI): http://indigenousedu.org/
[2] 'Imiloa Center in Hilo, HI: http://www.imiloahawaii.org
[3] I-WISE Conference website: http://iwiseconference.org/

**RECOMMENDATIONS:**

**1. Creating sufficient resources and funding to promote collaboration, research, and naming/renaming opportunities on telescope sites on indigenous lands, including those funded by NSF, NASA, and the DOE.** The power in a name, and in naming, is a belief common to nearly all indigenous peoples worldwide (see, e.g., Thornton 1997), as IK is deeply rooted in the traditional language base of each indigenous community. Recent examples of collaborative naming by nonindigenous and Native Hawaiian scholars through data taken by telescopes on Mauna Kea, Hawaii, include the interstellar asteroid 'Oumuamua (Witze 2019), and Pōwehi, the black hole depicted in the first image ever taken of one[4]. Other examples include the names given to a pair of unusual asteroids by Hawaiian immersion school students (Kamoʻoalewa and Kaʻepaokaʻāwela) that were recently adopted by the International Astronomical Union. See also the Astro2020 APC white paper by Kimura et al. on the naming of 'Oumuamua.

We urge astronomical facilities on indigenous lands to consider such collaboration avenues with local indigenous communities and youth education/research programs. Developing new contexts for indigenous languages to "live" in is key to continued progress towards the renormalization of IK while fostering close collaborations between IK and academic scientific inquiry. Astronomy could be a platform that affirms the living nature of indigenous languages and empowers the history of indigenous communities, while bringing deeper cultural meaning to our collective scientific progress.

**2. Greater funding should be allocated towards collaborative efforts that serve as successful models and roadmaps for how to coherently juxtapose indigenous and nonindigenous science.** Such partnerships, many of which are relatively new, have already had very positive results for everyone involved, moving beyond previous sterile isolating approaches and helping scientists, indigenous and nonindigenous, realize and reconnect with the richness and depth of their own and others' traditional knowledge. Examples of this "Collaboration with Integrity" (Maryboy, Begay & Peticolas 2012) pioneered by the Indigenous Education Institute, the 'Imiloa Astronomy Center in Hawai'i, and I-WISE (Indigenous Worldviews in Informal Science Education) include: Cosmic Serpent (Maryboy, Begay & Peticolas 2012), A Hua He Inoa[5], Envision MaunaKea[6], Native Universe[7], and MaunaKea Scholars[8]. These partnerships serve as

---

[4] https://www.nytimes.com/2019/04/13/science/powehi-black-hole.html
[5] https://imiloahawaii.org/news/a-hua-he-inoa-8e3ax
[6] http://www.envisionmaunakea.org/

roadmaps for new models of scientific collaboration that are worthy of agency investment (present and future), and long-term funding. Such collaboration is possible without the historical pattern of appropriation or assimilation into nonindigenous knowledge systems (Kimmerer 2015).

The other aspect to this recommendation is that false dichotomies must be discouraged in funded efforts. As members of the astronomical community, we must challenge ourselves to have the imagination to go beyond divisive dualities. Agencies and institutions should not create an artificial separation between the goals of western science and indigenous knowledge systems. Recent press coverage of the Thirty Meter Telescope in major news sources has sometimes referred to Hawaiians versus astronomers, spirituality/superstition/culture vs. science, glossing over the facts that Native Hawaiians have always been astronomers and that two-thirds of Hawaiians currently support scientific efforts on Mauna Kea that occur with Hawaiian community participation and input. From the examples stated earlier, emerging experiences from co-created scientific projects have the potential to reach new heights of progress. We can co-create this new reality with collective resolve and imagination, accepting the responsibility of establishing relationships through deep listening, and funding thoughtful dialogues and meetings involving indigenous voices (rather than the at-times troubled history of academic, anthropological or media representation of these voices).

**3.** Indigenous people have always been scientists, and have their own well-developed cosmologies integrated into worldviews based on generations of sky observations that have been used to serve their communities and foster well-being. IK embodies the principle that science is inseparable from the people doing it. Owing to several factors - the necessarily multigenerational nature of oral traditions, the time that it takes to conduct astronomical research in an indigenous context, conducting the observations to verify the oral tradition of elders - **we recommend that funding agencies support long-term, observation-based astronomy research conducted by indigenous knowledge holders to strengthen the intergenerational foundation of IK.** Such funding would ideally be administered and assessed on longer timescales than the usual 3 years to allow for the establishment of a strong network of relationships, a necessary precursor to establish trust among indigenous scientists/knowledge holders and non-indigenous collaborators. **We recommend 5-10 year grant timescales to allow the full development of proposed outcomes in an IK context.**

---

[7] http://www.nativeuniverse.org/
[8] https://maunakeascholars.com/

This may sound unfeasibly long, but observational astronomy has many areas requiring such long-term observations, such as astrometry or the 20 years of interdisciplinary collaboration that it took to make the first image of the black hole known by its Native Hawaiian name Pōwehi. One concern is that a longer grant timescale may also mean that funding rates would scale down proportionally and offer less grant funds to go around in an already super-competitive funding environment.  An additional concern is that the holistic nature of IK, which integrates many scientific disciplines, e.g. biology, ecology, medicine, astronomy, into a lived sustainable practice, means that an IK project does not easily fit into grant subfield categories, making it harder for indigenous scientists to procure sustained funding. **For all these reasons, we encourage dedicated agency funds for 5-10 year grants that invite indigenous scientific projects and research that allow for relationship/trust development and address long-term interdisciplinary science rooted in indigenous communities.**

**4. Funding agencies can lay the groundwork to recognize oral traditions as scientific references, on par with journal articles or other venues of peer-reviewed peer-confirmed scientific results.** Oral traditions, like nonindigenous scientific research, are integral systems of knowledge curation which are collaborative, build on previous knowledge and past results, and practice transmission across generations. One example from the past few decades has been the revitalization of Polynesian wayfinding and celestial navigation. This had not been practiced by Native Hawaiians for about 600 years. Had it not been for the oral tradition of Micronesian navigators, including master navigator Mau Piailug of Satawal, Micronesia who taught (non-instrument) oceanic and stellar navigation to Native Hawaiians in the 1970s[9], that knowledge would have been lost forever.

**5. Funding agencies can incentivize public and private organizations and institutions, including professional societies, to include the perspectives of indigenous astronomers and scientists in the rapid acceleration of human activity and presence in space.**  Dark skies are a human right, and especially important to the astronomy and cultural practices of indigenous peoples, navigators and wayfinding. We predict that by the decadal survey Astro2030, the issue of who space belongs to will be front and center, given the already rapid privatization of space. **We ask the funding agencies to help advocate for space being a resource that is held in community**

---

[9] http://www.hokulea.com/voyages/our-story/

**trust and belongs to everyone**, rather than the first-come first-serve approach that underlies the view that space belongs to no one. The recent slew of satellite launches from SpaceX, and the imminent onslaught of tens of thousands more to come from SpaceX and Amazon, imply a night sky that has reflecting satellites that greatly outnumber the stars visible to the human eye[10]. These decisions will reverberate for generations to come for astronomy[11] and all of humanity, especially indigenous cultural practices. Let us indigenize, not colonize, space.

**6. We ask the agencies to create a culturally-supported path for full participation of indigenous youth in science careers.** This could take a number of forms, including: contributing to the revitalization and inclusion of IK systems in mainstream science through collaboration with integrity (rather than assimilation or appropriation); funding conferences that help indigenous youth network and collaborate between individual tribes, peoples or countries; targeted early career efforts partnering academic institutions with underfunded tribal colleges; the use of ethnomathematics and ethnoscience; and, place-based education opportunities that naturally link math and science to culturally established ways of life. Through these venues, agencies will strengthen the motivation and incentives for indigenous youth to become a fully integrated part of a more vital inclusive scientific workforce for the future.

**Conclusion:**

We have shared a brief overview of astronomy's unique grounding in IK and relation to indigenous peoples. Given the history of the western hemisphere and the development of nonindigenous science, modern-day astronomy also has a unique obligation to indigenous lands and indigenous scientific practices. We have presented six actionable recommendations for the funding agencies that may be viewed as a charge to astronomy as a field to have better diversity/inclusion outcomes by Astro2030 related to IK and the adequate representation of Native and indigenous youth across a variety of astronomy career stages and institutions. These steps will move us collectively beyond merely improving the diversity of astronomy's future workforce towards the deeper long-term work of recognizing the true potential of IK: it has depth and breadth arising from multigenerational experiential scientific practices, and can lead to innovative solutions from all human ways of knowing that are truly needed for today's pressing global crises. Rather than asking future generations of astronomers to choose between science and culture, let us transform scientific culture.

---

[10] https://www.nytimes.com/2019/06/01/science/starlink-spacex-astronomers.html
[11] See position statement by the American Astronomical Society: https://aas.org/media/press-releases/aas-issues-position-statement-satellite-constellations

Last, we invite the readers of this white paper and representatives of the funding agencies to experience the wonderful outcomes already emerging from the approaches we have highlighted here by attending planned sessions on Collaboration with Integrity at the following meetings over the coming year: SACNAS 2019[12] (Oct. 31 - Nov. 2, 2019 in Honolulu, Hawaii) and the 235th meeting of the American Astronomical Society[13] (Jan. 4-8, 2020 in Honolulu, Hawaii).

---

[12] https://www.2019sacnas.org/
[13] https://aas.org/meetings/aas235